# Hierarchies in Dependence Logic[*]

Arnaud Durand [†]     Juha Kontinen[‡]

October 30, 2018

**Abstract**

We study fragments of dependence logic defined either by restricting the number $k$ of universal quantifiers or the width of dependence atoms in formulas. We find the sublogics of existential second-order logic corresponding to these fragments of dependence logic. We also show that these both ways of defining fragments of dependence logic give rise to a hierarchy in expressive power with respect to $k$.

## 1 Introduction

Dependence logic [19] extends first-order logic by dependence atomic formulas

$$=(x_1,\ldots,x_n) \tag{1}$$

the intuitive meaning of which is that the value of $x_n$ is completely determined by the values of $x_1,\ldots,x_{n-1}$. While it is known that dependence logic ($\mathscr{D}$) is equivalent to existential second-order logic (ESO) (and thus to NP over finite structures [4]) in expressive power, not much is known about the expressive power of its fragments. On the other hand, various fragments of ESO have been studied and the expressive power of the fragments of ESO is quite well understood (see, e.g., [1], [9], [18], [11], [3], and [8]). In this article we take the first steps towards charting the expressive power of fragments of dependence logic. The fragments studied in this article are defined by restricting the number of universal quantifiers or the width of the dependence atoms (the integer $n$ in (1)) in formulas. In both cases, we find exact subclasses of ESO corresponding to these fragments of dependence logic.

Let us briefly recall the history of dependence logic. In first-order logic the order of quantifiers determines the dependence relations between variables. For example, in the formula

$$\forall x_0 \exists x_1 \forall x_2 \exists x_3\, \phi,$$

the choice for $x_1$ depends on the value for $x_0$, and the choice for $x_3$ depends on the values of both universally quantified variables $x_0$ and $x_2$. Dependence logic generalizes the syntax of first-order logic to express dependencies between variables that are not first-order expressible. The first step in this direction was taken by Henkin [12] with his partially ordered quantifiers

$$\begin{pmatrix} \forall x_0 & \exists x_1 \\ \forall x_2 & \exists x_3 \end{pmatrix} \phi, \tag{2}$$

[*]The second author was supported by grant 127661 of the Academy of Finland.

[†]Institut de Mathématiques de Jussieu, Équipe de Logique, CNRS UMR 7586 - Université Paris Diderot, France. durand@logique.jussieu.fr

[‡]Department of Mathematics and Statistics, University of Helsinki, Finland. juha.kontinen@helsinki.fi

where $x_1$ depends only on $x_0$ and $x_3$ depends only on $x_2$. Walkoe [20] showed that every ESO-definable property can be defined by a sentence of the form (2) (the size of the quantifier prefix may vary) where $\phi$ is a quantifier-free first-order formula. Then Hintikka and Sandu [13] introduced Independence-Friendly (IF) logic extending first-order logic by so-called slashed quantifiers. For example, the formula (2) can be expressed in IF logic as

$$\forall x_0 \exists x_1 \forall x_2 \exists x_3 / \forall x_0 \phi,$$

where the quantifier $\exists x_3 / \forall x_0$ has the meaning that $x_3$ is "independent" of $x_0$ in the sense that a choice for the value of $x_3$ should not depend on the value of $x_0$. Dependence logic generalizes the approach of IF logic by detaching variable dependencies from the quantifiers and, instead, declaring them in terms of new atomic formulas. For example, the partially ordered quantifier (2) can be expressed in dependence logic as follows

$$\forall x_0 \exists x_1 \forall x_2 \exists x_3 (=(x_2, x_3) \wedge \phi).$$

It is known that dependence logic (also IF logic) is equivalent to ESO in expressive power. However, at the moment we do not have a good understanding of how various syntactic restrictions imposed on formulas of dependence logic reflect on their expressiveness and complexity. Some work has been done in this direction. In a recent doctoral thesis of Jarmo Kontinen [14] the complexity of open formulas of dependence logic was studied. The following result of [14] is in drastic contrast with classical logics. Define the formulas $\phi_1$ and $\phi_2$ as follows:

1. $\phi_1 := =(x,y) \vee =(u,v)$,

2. $\phi_2 := =(x,y) \vee =(u,v) \vee =(u,v)$.

Then the question of deciding whether a team $X$ satisfies $\phi_1$ is NL-complete and, for $\phi_2$, NP-complete. This result shows that already the quantifier-free part of dependence logic is as complex as the whole logic. Note also that the difference in the formulas $\phi_1$ and $\phi_2$ is that, in $\phi_2$, the disjunct $=(u,v)$ has two occurrences instead of one.

In this article we consider the following fragments $\mathscr{D}(k-\forall)$ and $\mathscr{D}(k-dep)$ of $\mathscr{D}$ defined by restricting the number of universal quantifiers or the width of dependence atoms in formulas, respectively. In other words, the fragment $\mathscr{D}(k-\forall)$ contains those formulas of $\mathscr{D}$ in which at most $k$ variables have been universally quantified and no reusing (i.e., requantification) of variables is allowed. On the other hand, in $\mathscr{D}(k-dep)$ we require that only dependence atoms $=(x_1,\ldots,x_n)$ satisfying $n \leq k+1$ may appear.

We show that $\mathscr{D}(k-dep)$ corresponds exactly to the fragment $\mathrm{ESO}_f(k\text{-ary})$ of ESO in which functions of arity at most $k$ are allowed to be quantified. Regarding $\mathscr{D}(k-\forall)$, we observe that this fragment is equivalent to the fragment $\mathrm{ESO}_f^1(k\forall, \exists^*)$ (see Definition 2.16) of ESO satisfying

$$\mathrm{ESO}_f^1(k\forall, \exists^*) \leq \mathrm{ESO}_f(k\text{-ary}, k\forall) \leq \mathrm{ESO}_f(k\text{-ary}),$$

where $\mathrm{ESO}_f(k\text{-ary}, k\forall)$ consists of those $\mathrm{ESO}_f(k\text{-ary})$-sentences that are in Skolem Normal Form and contain at most $k$ universal first-order quantifiers.

In the last section of this article we consider the expressive power of the logics $\mathscr{D}(k-\forall)$ and $\mathscr{D}(k-dep)$ over finite structures and investigate in particular whether, for varying $k$, these fragments form a strict hierarchy of expressivity. We recall a result of Ajtai [1] showing that in $\mathrm{ESO}_f(k\text{-ary})$ even cardinality of a $k+1$-ary relation cannot be expressed. Our results imply that the same holds for the logic $\mathscr{D}(k-dep)$ and that the hierarchy is strict when $k$ is less than the maximal arity of a relation in the signature. For the general case, we give some evidence that proving the strictness of the hierarchy may be hard since our results imply that this question is related to the Spectrum Arity Hierarchy Conjecture (see [6]).

On the other hand, in [11] Grandjean and Olive showed that, for any signature $\tau$,

$$\text{ESO}_f(k\forall) = \text{ESO}_f(k\text{-ary}, k\forall) = \text{NTIME}_{\text{RAM}}(n^k),$$

where $\text{NTIME}_{\text{RAM}}(n^k)$ denotes the family of classes of $\tau$-structures that can be recognizes by a non-deterministic Random Access Machine in time $O(n^k)$. The hierarchy theorem of Cook [2] implies that these classes form a strict hierarchy with respect to $k$ and hence the same holds for the logics $\text{ESO}_f(k\forall)$. We show that

$$\text{ESO}_f(k\forall) \leq \mathscr{D}(2k - \forall) \leq \text{ESO}_f(2k\forall),$$

implying a hierarchy theorem for the logics $\mathscr{D}(k - \forall)$.

This article is organized as follows. In Section 2, we review some basic properties and results regarding dependence logic. We also recall some relevant results characterizing subclasses of NP in terms of fragments of ESO. Section 3 contains our main results connecting fragments of dependence logic with that of ESO. Finally, in Section 4, we use our results to identify expressibility hierarchies within dependence logic.

## 2 Preliminaries

In this section we first define dependence logic and recall some basic results on it. Then we review results in computational complexity and descriptive complexity theory that will be needed.

### 2.1 Dependence Logic

We begin with the syntax of dependence logic.

**Definition 2.1** ([19])**.** The syntax of $\mathscr{D}$ extends the syntax of FO, defined in terms of $\vee, \wedge, \neg, \exists$ and $\forall$, by atomic dependence formulas of the form

$$=(t_1, \ldots, t_n), \tag{3}$$

where $t_1, \ldots, t_n$ are terms. For a signature $\tau$, $\mathscr{D}[\tau]$ denotes the set of $\tau$-formulas of $\mathscr{D}$.

The meaning of the dependence formula (3) is that the value of the term $t_n$ is functionally determined by the values of the terms $t_1, \ldots, t_{n-1}$. Hence the meaning of the formula $=(t)$ is that the value of the term $t$ depends on nothing, i.e., is constant. As a singular case we have $=()$, which we take to be universally true.

**Definition 2.2.** The set $\text{Fr}(\phi)$ of free variables of a formula $\phi \in \mathscr{D}$ is defined as for first-order logic, except that we have the new case

$$\text{Fr}(=(t_1, \ldots, t_n)) = \text{Var}(t_1) \cup \cdots \cup \text{Var}(t_n),$$

where $\text{Var}(t_i)$ is the set of variables occurring in term $t_i$. If $\text{Fr}(\phi) = \emptyset$, we call $\phi$ a sentence.

The semantics of $\mathscr{D}$ is formulated using the concept of a *Team*. Let $\mathfrak{A}$ be a model with domain $A$. *Assignments* of $\mathfrak{A}$ are finite mappings from variables into $A$. The value of a term $t$ in an assignment $s$ is denoted by $t^{\mathfrak{A}}\langle s \rangle$. If $s$ is an assignment, $x$ a variable, and $a \in A$, then $s(a/x)$ denotes the assignment (with domain $\text{dom}(s) \cup \{x\}$) that agrees with $s$ everywhere except that it maps $x$ to $a$. For an assignment $s$, and a tuple of variables $\overline{x} = (x_1, \ldots, x_n)$, we sometimes denote the tuple $(s(x_1), \ldots, s(x_n))$ by $s(\overline{x})$.

**Definition 2.3.** Let $A$ be a set and $\{x_1, \ldots, x_k\}$ a finite (possibly empty) set of variables. A *team* $X$ of $A$ with domain $\text{dom}(X) = \{x_1, \ldots, x_k\}$ is any set of assignments from the variables $\{x_1, \ldots, x_k\}$ into the set $A$.

We denote by $rel(X)$ the $k$-ary relation of $A$ corresponding to $X$

$$rel(X) = \{(s(x_1),\ldots,s(x_k)) : s \in X\}.$$

If $X$ is a team of $A$, and $F\colon X \to A$, we use $X(F/x_n)$ to denote the team $\{s(F(s)/x_n) : s \in X\}$ and $X(A/x_n)$ the team $\{s(a/x_n) : s \in X \text{ and } a \in A\}$.

We are now ready to define the semantics of dependence logic. Signatures $\tau$ are assumed to be finite and they may contain constants, relations and function symbols. In this article we consider only formulas in negation normal form (NNF), i.e., negation is allowed to appear only in front of atomic formulas. This is not a restriction since any formula of dependence logic can be transformed into negation normal form [19]. Atomic formulas and their negations are called literals.

**Definition 2.4** ([19]). Let $\mathfrak{A}$ be a model and $X$ a team of $A$. The satisfaction relation $\mathfrak{A} \models_X \phi$ is defined as follows:

1. If $\phi$ is a first-order literal, then $\mathfrak{A} \models_X \phi$ iff for all $s \in X$: $\mathfrak{A} \models_s \phi$.

2. $\mathfrak{A} \models_X =(t_1,\ldots,t_n)$ iff for all $s, s' \in X$ such that $t_1^{\mathfrak{A}}\langle s \rangle = t_1^{\mathfrak{A}}\langle s' \rangle, \ldots, t_{n-1}^{\mathfrak{A}}\langle s \rangle = t_{n-1}^{\mathfrak{A}}\langle s' \rangle$, we have $t_n^{\mathfrak{A}}\langle s \rangle = t_n^{\mathfrak{A}}\langle s' \rangle$.

3. $\mathfrak{A} \models_X \neg =(t_1,\ldots,t_n)$ iff $X = \emptyset$.

4. $\mathfrak{A} \models_X \psi \wedge \phi$ iff $\mathfrak{A} \models_X \psi$ and $\mathfrak{A} \models_X \phi$.

5. $\mathfrak{A} \models_X \psi \vee \phi$ iff $X = Y \cup Z$ such that $\mathfrak{A} \models_Y \psi$ and $\mathfrak{A} \models_Z \phi$.

6. $\mathfrak{A} \models_X \exists x_n \psi$ iff $\mathfrak{A} \models_{X(F/x_n)} \psi$ for some $F\colon X \to A$.

7. $\mathfrak{A} \models_X \forall x_n \psi$ iff $\mathfrak{A} \models_{X(A/x_n)} \psi$.

Above, we assume that the domain of $X$ contains the variables free in $\phi$. Finally, a sentence $\phi$ is true in a model $\mathfrak{A}$ ($\mathfrak{A} \models \phi$) if $\mathfrak{A} \models_{\{\emptyset\}} \phi$.

Next we define the concepts of logical consequence and equivalence for formulas of dependence logic.

**Definition 2.5.** Let $\phi$ and $\psi$ be formulas of dependence logic. The formula $\psi$ is a *logical consequence* of $\phi$,

$$\phi \Rightarrow \psi,$$

if for all models $\mathfrak{A}$ and teams $X$, with $\text{Fr}(\phi) \cup \text{Fr}(\psi) \subseteq \text{dom}(X)$, and $\mathfrak{A} \models_X \phi$ we have $\mathfrak{A} \models_X \psi$. The formulas $\phi$ and $\psi$ are *logically equivalent*,

$$\phi \equiv \psi,$$

if $\phi \Rightarrow \psi$ and $\psi \Rightarrow \phi$.

## 2.2 Basic properties of dependence logic

In this section we recall some basic properties of dependence logic.

Let $X$ be a team with domain $\{x_1,\ldots,x_k\}$ and $V \subseteq \{x_1,\ldots,x_k\}$. Denote by $X \upharpoonright V$ the team $\{s \upharpoonright V : s \in X\}$ with domain $V$. The following lemma shows that the truth of a formula depends only on the interpretations of the variables occurring free in the formula.

**Lemma 2.6** ([19]). *Suppose $V \supseteq \mathrm{Fr}(\phi)$. Then $\mathfrak{A} \models_X \phi$ if and only if $\mathfrak{A} \models_{X \upharpoonright V} \phi$.*

The following fact is also a very basic property of all formulas of dependence logic:

**Proposition 2.7** ([19]). *Let $\phi$ be a formula of dependence logic, $\mathfrak{A}$ a model, and $Y \subseteq X$ teams. Then $\mathfrak{A} \models_X \phi$ implies $\mathfrak{A} \models_Y \phi$.*

On the other hand, the expressive power of sentences of $\mathscr{D}$ coincides with that of ESO:

**Theorem 2.8** ([19]). *$\mathscr{D} = \mathrm{ESO}$.*

Theorem 2.8 does not tell us anything about formulas of dependence logic with free variables. An upperbound for the complexity of formulas of $\mathscr{D}$ is provided by the following result.

**Theorem 2.9** ([19]). *Let $\tau$ be a signature and $\phi$ a $\mathscr{D}[\tau]$-formula with free variables $x_1, \ldots, x_k$. Then there is a $\tau \cup \{R\}$-sentence $\psi$ of ESO, in which $R$ appears only negatively, such that for all models $\mathfrak{A}$ and teams $X$ with domain $\{x_1, \ldots, x_k\}$:*
$$\mathfrak{A} \models_X \phi \iff (\mathfrak{A}, rel(X)) \models \psi.$$

In [15] it was shown that also the converse holds.

**Theorem 2.10** ([15]). *Let $\tau$ be a signature and $R$ a $k$-ary relation symbol such that $R \notin \tau$. Then for every $\tau \cup \{R\}$-sentence $\psi$ of ESO, in which $R$ appears only negatively, there is a $\tau$-formula $\phi$ of dependence logic with free variables $x_1, \ldots, x_k$ such that, for all $\mathfrak{A}$ and $X$ with domain $\{x_1, \ldots, x_k\}$:*
$$\mathfrak{A} \models_X \phi \iff (\mathfrak{A}, rel(X)) \models \psi \vee \forall \bar{y} \neg R(\bar{y}). \tag{4}$$

*Proof.* The disjunct $\forall \bar{y} \neg R(\bar{y})$ is needed on the right because the empty team $X = \emptyset$ satisfies all formulas of dependence logic but $\psi$ need not always be true in the case $rel(X) = \emptyset$ [16]. □

Theorem 2.10 shows that formulas of dependence logic correspond in a precise way to the negative fragment of ESO and are therefore very expressive. Furthermore, the results of [14] discussed in the Introduction show that already for certain quantifier-free formulas $\phi \in \mathscr{D}$, the corresponding sentence $\psi \in \mathrm{ESO}$ (see Theorem 2.9) defines a NP-complete problem. On the other hand, if we restrict attention to formulas that do not contain dependence atomic formulas as subformulas, we lose much of the expressive power.

**Definition 2.11.** A formula $\phi$ of $\mathscr{D}$ is called a first-order formula if it does not contain dependence atomic formulas as subformulas.

**Theorem 2.12.** *Let $\phi$ be a first-order formula of dependence logic. Then for all $\mathfrak{A}$ and $X$:*

1. $\mathfrak{A} \models_{\{s\}} \phi \iff \mathfrak{A} \models_s \phi$.

2. $\mathfrak{A} \models_X \phi \iff$ *for all $s \in X$: $\mathfrak{A} \models_s \phi$.*

The following proposition shows that both the existential fragment of $\mathscr{D}$, and the fragment allowing only dependence atoms of width 1 (i.e., dependence atoms $=(t_1)$), collapse also to first-order logic.

**Proposition 2.13.** *Suppose that a sentence $\phi \in \mathscr{D}$ satisfies either of the following*

1. *$\phi$ is in NNF and does not contain universal quantifiers,*

2. *$\phi$ contains only dependence atoms of width 1 as subformulas.*

*Then $\phi$ is equivalent to a first-order sentence.*

*Proof.* The result of Case 2 has been shown by Galliani in [7]. He studies the fragment of $\mathscr{D}$ allowing only dependence atoms of width 1 and shows (independently of this article) that, from sentences, they can be eliminated using existential quantification in a similar fashion to Lemma 3.2.

Case 1 is proved using induction on $\phi$. It is straightforward to show that for all $\mathfrak{A}$ and assignments $s$:
$$\mathfrak{A} \models_{\{s\}} \phi \iff \mathfrak{A} \models_s \phi^*,$$
where $\phi^*$ is obtained from $\phi$ by replacing dependence atoms in terms of $\top$. □

In the next lemma, we list certain properties of dependence logic that will be used later.

**Lemma 2.14.** *Formulas of dependence logic satisfy the following properties:*

1. $\exists x(\phi \vee \psi) \equiv \exists x\phi \vee \exists x\psi$,

2. $\exists x(\phi \wedge \psi) \equiv \exists x\phi \wedge \psi$, *if $x$ is not free in $\psi$,*

3. $\forall x(\phi \wedge \psi) \equiv \forall x\phi \wedge \forall x\psi$,

4. $\forall x(\phi \vee \psi) \equiv \forall x\phi \vee \psi$, *if $x$ is not free in $\psi$,*

5. *Every formula of dependence logic can be transformed into prenex normal form.*

6. *The meaning of a formula $\phi$ is invariant under replacing a subformula $\psi$ of $\phi$ by $\psi'$ such that $\psi \equiv \psi'$ and $\neg\psi \equiv \neg\psi'$.*

*Proof.* For 1 and 3 see Lemma 3.23, for 2 and 4 see Exercise 3.49, and for 5 see Exercise 3.51 in [19]. Finally, we note that 6 is based on the strong compositionality of dependence logic (see Lemma 3.25 in [19] for the exact formulation). The assumption $\neg\psi \equiv \neg\psi'$ will not be relevant for our purposes. □

We end this section by defining the fragments of dependence logic and ESO that will be discussed in the following sections.

**Definition 2.15.** Let $k \in \mathbb{N}^*$.

- Denote by $\mathscr{D}(k - \forall)$ the class of NNF sentences $\phi$ of $\mathscr{D}$ whose every variable is quantified exactly once (no reusing of variables), and $\phi$ contains at most $k$ occurrences of the quantifier $\forall$.

- Denote by $\mathscr{D}(k - dep)$ the class of NNF sentences of $\mathscr{D}$ in which dependence atoms of width at most $k+1$ (i.e., atoms of the form $=(t_1, \ldots, t_l)$, where $l \leq k+1$) may appear.

- Denote by ESO($k$-ary) the class of ESO-sentences
$$\exists X_1 \ldots \exists X_n \psi,$$
in which the relation symbols $X_i$ are at most $k$-ary and $\psi$ is a first-order formula.

- Denote by $\text{ESO}_f(k$-ary) the class of ESO-sentences
$$\exists f_1 \ldots \exists f_n \psi,$$
in which the function symbols $f_i$ are at most $k$-ary and $\psi$ is a first-order formula.

- Denote by $\mathrm{ESO}_f(k\text{-ary}, m\forall)$ the class of ESO-sentences in Skolem Normal Form

$$\exists f_1 \ldots \exists f_n \forall x_1 \ldots \forall x_r \psi,$$

  in which the function symbols $f_i$ are at most $k$-ary and $r \leq m$.

- Denote by $\mathrm{ESO}_f(m\forall)$ the class of ESO-sentences in Skolem Normal Form

$$\exists f_1 \ldots \exists f_n \forall x_1 \ldots \forall x_r \psi,$$

  where $r \leq m$.

By abuse of terminology, we identify these classes of sentences with the classes of properties they define.

Finally we define some further fragments of ESO that will play a central rôle in the results of Section 3.1.

**Definition 2.16.** Let $k \in \mathbb{N}^*$.

1. Denote by $\mathrm{ESO}_f^1(k\text{-ary})$ the class of ESO-sentences

$$\exists f_1 \ldots \exists f_n \psi,$$

   in which each function symbol $f_i$ is at most $k$-ary and there exists $i_1, \ldots, i_m$, pairwise distinct, such that all terms and subterms in $\psi$ with $f_i$ as the outermost symbol are of the form $f_i(x_{i_1}, \ldots, x_{i_m})$.

2. Denote by $\mathrm{ESO}_f^1(k\forall)$ the class of ESO-sentences in Skolem Normal Form

$$\exists f_1 \ldots \exists f_n \forall x_1 \ldots \forall x_p \psi,$$

   such that $p \leq k$ and in which, for each symbol $f_i$, there exists $i_1, \ldots, i_m$, pairwise distinct, such that all terms and subterms in $\psi$ with $f_i$ as the outermost symbol are of the form $f_i(x_{i_1}, \ldots, x_{i_m})$.

3. Denote by $\mathrm{ESO}_f^1(k\forall, \exists^*)$ the class of ESO-sentences of the form

$$\exists f_1 \ldots \exists f_n Q^1 x_1 \ldots Q^p x_p \psi,$$

   where $Q^i \in \{\exists, \forall\}$, and the number of $i$, for $1 \leq i \leq p$, such that $Q^i = \forall$ is at most $k$. Furthermore, for each symbol $f_i$ there must exists $x_{i_1}, \ldots, x_{i_m}$, pairwise distinct, such that all terms and subterms in $\psi$ with $f_i$ as the outermost symbol are of the form $f_i(x_{i_1}, \ldots, x_{i_m})$.

We identify these classes of sentences with the classes of properties they define.

It is worth noting that the definition of the logic $\mathrm{ESO}_f^1(k\forall)$ forces the functions $f_i$ to be at most $k$-ary, whereas in $\mathrm{ESO}_f^1(k\forall, \exists^*)$, the functions $f_i$ can have arity greater than $k$.

## 2.3 Background in complexity

In this section we review some concepts and results in complexity theory and descriptive complexity theory. We assume that the reader is familiar with the basics of computational complexity theory.

Descriptive complexity theory studies and applies logical methods in the area of computational complexity theory. The seminal result in the field was Fagin's [4] characterization of NP in terms of problems describable in ESO. Since then, most of the central complexity classes have been given such logical characterization. Fagin's characterization of NP implies, by Theorem 2.8, that

$$\mathscr{D} = \mathrm{NP},$$

i.e., for every signature $\tau$, and every class $K \subseteq \mathrm{Str}(\tau)$ of finite structures: $K = \mathrm{Mod}(\phi)$ for some $\phi \in \mathscr{D}[\tau]$ iff $L_K \in \mathrm{NP}$, where $L_K \subseteq \{0,1\}^*$ is a language encoding the class $K$.

In this paper we are interested in fragments of dependence logic and NP. We denote by $\mathrm{NTIME}(n^k)$ the class of languages that can be recognized by some nondeterministic Turing Machine in time $O(n^k)$. Lynch [17] observed that the exponent $k$ in $\mathrm{NTIME}(n^k)$ corresponds roughly to the arity of relations quantified in formulas of ESO:

**Theorem 2.17.** *If $L \in \mathrm{NTIME}(n^k)$ then there is a sentence $\phi \in \mathrm{ESO}(k\text{-}ary)(s,+)$ that defines the class of string structures that corresponds to L, where s and $+$ are built-in relations for successor and addition. Furthermore, if $k \geq 2$, then $+$ is not needed.*

In Theorem 2.17, the first-order part of $\phi$ has the quantifier prefix $\forall^* \exists^*$. In [10, 18] and later in [11], Grandjean and Olive showed that, when considering Random Access Machine as the computation model, a tighter correspondence can be proved and, actually, an exact characterization of fragments of NP on RAM's can be obtained.

Grandjean and Olive consider $\tau$-NRAM's, a nondeterministic RAM that takes an arbitrary $\tau$-structure as input (see [11] for a complete description of this model). A problem $L$ on $\tau$-structures is in the class $\mathrm{NTIME}_{\mathrm{RAM}}(n^k)$, $k \in \mathbb{N}^*$, if there exists a $\tau$-NRAM $M$ that recognizes every structure of $L$ and such that: each computation of $M$ on a structure $\mathfrak{A}$ with domain $A$ of size $n$ uses only integers in $O(n^k)$ (for address or register contents) and stops after $O(n^k)$ steps. To count the cost of the computation, the uniform cost measure is adopted.

Olive proved the following result for signatures consisting of unary functions [18]. It was later generalized for any kind of input structures in [11].

**Theorem 2.18.** *Let $k \in \mathbb{N}^*$ and let $\tau$ be any signature. Over $\tau$-structures:*

$$\mathrm{ESO}_f(k\text{-}ary, k\forall) = \mathrm{ESO}_f(k\forall) = \mathrm{NTIME}_{\mathrm{RAM}}(n^k).$$

It is worth noting that, even for $k = 1$, no built-in relations need to assumed in the above theorem (the result of [18] used built-in relations in the case $k = 1$). Note that $n$ is the domain size. Hence, if the maximal arity in $\tau$ is greater than $k$, then the number of steps in a $O(n^k)$ computation is less than the input size. Let us now recall the hierarchy theorem for nondeterministic time by Cook [2].

**Theorem 2.19.** *for every $k \in \mathbb{N}^*$: $\mathrm{NTIME}_{\mathrm{RAM}}(n^k) < \mathrm{NTIME}_{\mathrm{RAM}}(n^{k+1})$.*

The lemma below shows that the classical hierarchy result also applies to the "sublinear" case.

**Lemma 2.20.** *For all $h, k \in \mathbb{N}^*$ with $k < h$ and for every signature $\tau$ of arity $h$, it holds:*

$$\mathrm{ESO}_f(k\forall)[\tau] < \mathrm{ESO}_f((k+1)\forall)[\tau].$$

*Proof.* We prove the result in the hardest case of a signature $\tau$ restricted to one function symbol only (this implies the separation for all richer signatures). Let $\mathscr{P}$ be the set of $h$-ary structures $\mathfrak{A}$ on domain $A$ (identified with $\{0, \ldots, n-1\}$) over one function $F$ and one constant $0$ defined by:

$$\forall a_1 \in A \ldots \forall a_{k+1} \in A \; F(a_1, \ldots, a_{k+1}, 0, \ldots, 0) = 0$$

and arbitrary otherwise. In some sense, the functions $F$ in structures of $\mathscr{P}$ are constant on $A^{k+1}$ values which are chosen for convenience on the first $k+1$ coordinates. Clearly, $\mathscr{P} \in \mathrm{ESO}_f((k+1)\forall)$. Suppose now that $\mathscr{P}$ is definable in $\mathrm{ESO}_f(k\forall)$ by a formula $\Phi$ below:

$$\exists f_1 \ldots \exists f_t \forall x_1 \ldots \forall x_k \varphi,$$

where the $f_i$ are functions (we do not even need to suppose they are $k$-ary). Let $n \in \mathbb{N}$ and let $\mathfrak{A} \in \mathscr{P}$ on domain $A$ with $|A| = n$. By hypothesis, $\mathfrak{A} \models \Phi$. Then, there exists $f_1, \ldots, f_t$ on $A$ such that:

$$(\mathfrak{A}, f_1, \ldots, f_t) \models \forall x_1 \ldots \forall x_k \varphi.$$

Let us consider the propositional constraint $\phi$:

$$\bigwedge_{a_1=0}^{n-1} \ldots \bigwedge_{a_k=0}^{n-1} \varphi(a_i/x_i).$$

The formula $\phi$ is of size $O(n^k)$. Let now $L$ be the set of $h$-ary tuples $(a_{i_1}, \ldots, a_{i_h})$ such that $F(a_{i_1}, \ldots, a_{i_h})$ appears in formula $\varphi(a_i/x_i)$ (after substitution of all nested terms by their values which is possible since the interpretations of the $f_i$s are also known). The set $L$ is of size bounded by $cn^k$ for some constant $c$ depending on $|\varphi|$. Then, if $n$ is big enough, at least one tuple $(b_1, \ldots, b_{k+1}, 0, \ldots, 0) \in A^h$ does not belong to $L$. Now form a new structure $\mathfrak{A}'$ over the same domain $A$ with a new function $F'$ equal to $F$ on all elements of $A^h$ except on $(b_1, \ldots, b_{k+1}, 0, \ldots, 0)$ where the following holds:

$$F'(b_1, \ldots, b_{k+1}, 0, \ldots, 0) \neq 0.$$

Clearly, since $(b_1, \ldots, b_{k+1}, 0, \ldots, 0) \notin L$ then $(\mathfrak{A}', f_1, \ldots, f_t)$ satisfies the constraint

$$\bigwedge_{a_1=0}^{n-1} \ldots \bigwedge_{a_k=0}^{n-1} \varphi(a_i/x_i).$$

Hence,

$$(\mathfrak{A}', f_1, \ldots, f_t) \models \forall x_1 \ldots \forall x_k \varphi.$$

and then $\mathfrak{A}' \models \Phi$. But, since $F'(b_1, \ldots, b_{k+1}, 0, \ldots, 0) \neq 0$, then $\mathfrak{A}' \notin \mathscr{P}$. This contradicts our assumption that $\mathscr{P}$ is definable by $\Phi$. □

Hence, this last result, together with Theorem 2.19 implies the following corollary.

**Corollary 2.21.** *Let $\tau$ be any signature and $k \in \mathbb{N}^*$. On $\tau$-structures, $\mathrm{ESO}_f(k\forall) < \mathrm{ESO}_f((k+1)\forall)$.*

# 3 Relations between fragments of $\mathscr{D}$ and ESO

In this section we study the relations between fragments of dependence logic and ESO. We consider first the logics $\mathscr{D}(k-dep)$. The case $k = 0$ is solved by Proposition 2.13, hence we assume that $k \geq 1$.

The following proposition gives us a direct correspondence between certain sentences of $\mathscr{D}$ and ESO. This correspondence is slightly more general compared to that of Theorem 6.15 [19].

**Proposition 3.1.** *Let $\phi \in \mathscr{D}$ be a sentence of the form*

$$Q^1 x_1 \ldots Q^m x_m \exists y_1 \ldots \exists y_n ( \bigwedge_{1 \leq j \leq n} {=}(\bar{z}^j, y_j) \wedge \theta),$$

*where $Q^i \in \{\exists, \forall\}$, all the quantified variables are pairwise distinct, $\theta$ is a quantifier-free first-order formula, and each variable in $\bar{z}^j$ is in $\{x_1, \ldots, x_m\}$ and they are also pairwise distinct. Then $\phi$ is equivalent to the ESO-sentence $\chi$*

$$\exists f_1 \ldots \exists f_n Q^1 x_1 \ldots Q^m x_m \theta',$$

*where $\theta'$ is obtained from $\theta$ by replacing every occurrence of $y_i$ by the term $f_i(\bar{z}^i)$. Conversely, let $\chi$ be an ESO-sentence*

$$\exists f_1 \ldots \exists f_n Q^1 x_1 \ldots Q^m x_m \theta',$$

*in which, for each symbol $f_i$, there exists $i_1, \ldots, i_p$, pairwise distinct, such that all terms and subterms in $\theta'$ with $f_i$ as the outermost symbol are of the form $f_i(x_{i_1}, \ldots, x_{i_p})$, then $\chi$ is equivalent to a sentence $\phi \in \mathscr{D}$ as above.*

*Proof.* We will show that the sentences $\phi$ and $\chi$ are equivalent. Let $\mathfrak{A}$ be arbitrary and suppose that

$$\mathfrak{A} \models \phi.$$

This implies that there is a team $X$ such that

$$\mathfrak{A} \models_X \exists y_1 \ldots \exists y_n ( \bigwedge_{1 \leq j \leq n} {=}(\bar{z}^j, y_j) \wedge \theta), \qquad (5)$$

where $X$ is constructed by evaluating the quantifier prefix $Q^1 x_1 \ldots Q^m x_m$. Furthermore, (5) implies that there are functions $F_i : X_{i-1} \to A$, for $1 \leq i \leq n$, such that

$$\mathfrak{A} \models_{X_n} ( \bigwedge_{1 \leq j \leq n} {=}(\bar{z}^j, y_j) \wedge \theta), \qquad (6)$$

where $X_0 = X$, and $X_i = X_{i-1}(F_i/y_i)$. By the first conjunct in (6), the values $F_i(s)$ of $F_i$ are determined by the values $s$ assigns to the variables in $\bar{z}^i$. We can now choose functions $g_i : A^{|\bar{z}^i|} \to A$ satisfying $g_i(\bar{a}) = F_i(s)$ for all $\bar{a}$ such that $\bar{a} = s(\bar{z}^i)$ for some $s \in X_{i-1}$. We will next show that

$$(\mathfrak{A}, g_1, \ldots, g_n) \models_X \theta'. \qquad (7)$$

Recall that $\theta'$ is a first-order formula of dependence logic and hence, by Theorem 2.12, (7) holds iff $(\mathfrak{A}, g_1, \ldots, g_n) \models_s \theta'$ for each $s \in X$. We can now show, using induction on the construction of $\theta$, that for all $s \in X_n$ (recall that $\mathrm{dom}(s) = \{x_1, \ldots, x_m, y_1, \ldots, y_n\}$) it holds that

$$\mathfrak{A} \models_s \theta \iff (\mathfrak{A}, g_1, \ldots, g_n) \models_{s'} \theta', \qquad (8)$$

where $s' = s \upharpoonright \{x_1, \ldots, x_m\}$. The key to this result is the fact that, for $1 \leq i \leq n$, the interpretation of the variable $y_i$ and the term $f_i(\overline{z}^i)$ agree:

$$s(y_i) = F_i(s \upharpoonright \{x_1, \ldots, x_m, y_1, \ldots, y_{i-1}\}) = g_i(s'(\overline{z}^i)) = f_i(\overline{z}^i)^{(\mathfrak{A}, g_1, \ldots, g_n)} \langle s' \rangle.$$

This implies, for any complex term $t(y_1, \ldots, y_n)$, that the interpretations of the terms $t(y_1, \ldots, y_n)$ and

$$t(f_1(\overline{z}^1)/y_1, \ldots, f_n(\overline{z}^n)/y_n)$$

agree for $s$ and $s'$, respectively. With these observations, the induction in (8) is straightforward.

Now, by (6) and Proposition 2.7, for all $s \in X_n$, it holds that

$$\mathfrak{A} \models_s \theta.$$

Hence, by (8) and Theorem 2.12 again, we get that

$$(\mathfrak{A}, g_1, \ldots, g_n) \models_X \theta'.$$

This implies

$$(\mathfrak{A}, g_1, \ldots, g_n) \models Q^1 x_1 \ldots Q^m x_m \theta',$$

and, finally, that

$$\mathfrak{A} \models \exists f_1 \ldots \exists f_n Q^1 x_1 \ldots Q^m x_m \theta'.$$

The implication "$\chi \Rightarrow \phi$" follows by essentially reversing the steps above.

Remark that by construction, for each symbol $f_i$, there exists $i_1, \ldots, i_p$, pairwise distinct, such that all terms and subterms with $f_i$ as the outermost symbol in $\chi$ are of the form $f_i(x_{i_1}, \ldots, x_{i_p})$. Hence, the reciprocal result follows analogously. $\square$

Our goal is to find a subclass of ESO corresponding to $\mathscr{D}(k - dep)$. We will use Proposition 3.1 to achieve this. The following lemma allows us to transform a sentence in prenex normal form to the form required in Proposition 3.1.

**Lemma 3.2.** *Let $\psi \in \mathscr{D}$ be a quantifier-free formula whose dependence atomic subformulas are of the form $=(z_1, \ldots, z_m)$ for some pairwise distinct variables $z_1, \ldots, z_m$. Then $\psi$ is equivalent to a formula of the form*

$$\exists y_1 \ldots \exists y_n (\bigwedge_{1 \leq j \leq n} =(\overline{z}^j, y_j) \wedge \theta),$$

*where*

1. *$\theta$ is a quantifier-free formula without dependence atoms,*

2. *$y_i$ does not appear in $\overline{z}^j$ for $1 \leq j \leq n$,*

3. *the number and the width of the dependence atoms in the first conjunct corresponds to that of $\psi$.*

*Proof.* The claim is proved using induction on the formula $\psi$. If $\psi$ is a first-order literal then the claim holds trivially. If $\psi$ is of the form $=(z_1, \ldots, z_m)$, then we can transform $\psi$ into the following equivalent form satisfying the claim:

$$\exists y_1 (=(z_1, \ldots, z_{m-1}, y_1) \wedge y_1 = z_m).$$

Note that we do not have to consider the case of $\neg = (z_1, \ldots, z_m)$ since this formula is equivalent to $\bot$. Assume then that $\psi := \phi_1 \vee \phi_2$. By the induction hypothesis

$$\phi_1 \equiv \exists y_1 \ldots \exists y_n ( \bigwedge_{1 \leq j \leq n} =(\bar{z}^j, y_j) \wedge \theta_1), \tag{9}$$

and

$$\phi_2 \equiv \exists y_{n+1} \ldots \exists y_{n+m} ( \bigwedge_{n+1 \leq j \leq n+m} =(\bar{z}^j, y_j) \wedge \theta_2). \tag{10}$$

We may assume that the variables $y_1, \ldots, y_n$ and $y_{n+1}, \ldots, y_{n+m}$ do not appear in the formulas (10) and (9), respectively. Therefore, by Lemma 2.6 it follows that

$$\phi_2 \equiv \exists y_1 \ldots \exists y_n \phi_2,$$

and

$$\bigwedge_{1 \leq j \leq n} =(\bar{z}^j, y_j) \wedge \theta_1 \equiv \exists y_{n+1} \ldots \exists y_{n+m} ( \bigwedge_{1 \leq j \leq n} =(\bar{z}^j, y_j) \wedge \theta_1).$$

Using the above equivalences, we may now use successively clause 1 of Lemma 2.14 to show that $\psi$ is equivalent to the formula $\psi^*$

$$\psi^* := \exists y_1 \ldots \exists y_n \exists y_{n+1} \ldots \exists y_{n+m} (( \bigwedge_{1 \leq j \leq n} =(\bar{z}^j, y_j) \wedge \theta_1) \vee ( \bigwedge_{n+1 \leq j \leq n+m} =(\bar{z}^j, y_j) \wedge \theta_2)).$$

Next we will show how to transform the quantifier-free part $\chi$ of $\psi^*$ into the desired form

$$\chi := (( \bigwedge_{1 \leq j \leq n} =(\bar{z}^j, y_j) \wedge \theta_1) \vee ( \bigwedge_{n+1 \leq j \leq n+m} =(\bar{z}^j, y_j) \wedge \theta_2)).$$

Let $\mathfrak{A}$ be a model and $X$ a team whose domain consists of the free variables of $\psi^*$. Suppose $\mathfrak{A} \models_X \psi^*$. Then, there are functions $F_i$, for $1 \leq i \leq n+m$, such that $\mathfrak{A} \models_{X^*} \chi$, where $X^* = X(F_1/y_1) \cdots (F_{n+m}/y_{n+m})$. By the semantics of disjunction, there are teams $Y, Z$ such that $X^* = Y \cup Z$ and

$$\mathfrak{A} \models_Y ( \bigwedge_{1 \leq j \leq n} =(\bar{z}^j, y_j) \wedge \theta_1) \tag{11}$$

and

$$\mathfrak{A} \models_Z ( \bigwedge_{n+1 \leq j \leq n+m} =(\bar{z}^j, y_j) \wedge \theta_2).$$

Since the variables $y_{n+1}, \ldots, y_{n+m}$ do not appear in the formula in (11), it is obvious that we can modify the values of these variables in $Y$ to get $Y'$ such that $\mathfrak{A} \models_{Y'} \theta_1$, and

$$\mathfrak{A} \models_{Y' \cup Z} \bigwedge_{n+1 \leq j \leq n+m} =(\bar{z}^j, y_j).$$

More precisely, we do the following (note that we are modifying the values of the functions $F_{n+1}, \ldots, F_{n+m}$ in $Y$) for $n+1 \leq i \leq n+m$:

- For all $s \in Y$ such that for no $s' \in Z$, $s(\bar{z}^i) = s'(\bar{z}^i)$, we set $s(y_i) = a$, for some fixed $a \in A$.
- For all $s \in Y$ such that $s(\bar{z}^i) = s'(\bar{z}^i)$ for some $s' \in Z$, we set $s(y_i) = s'(y_i)$.

Analogously, by modifying the values of the variables $y_1,\ldots,y_n$ in $Z$, we find $Z'$ such that $\mathfrak{A}\models_{Z'}\theta_2$, and

$$\mathfrak{A}\models_{Y\cup Z'}\bigwedge_{1\leq j\leq n}=(\bar{z}^j,y_j).$$

Since $Y$ and $Y'$ and $Z$ and $Z'$ only differ on the values of the variables $y_{n+1},\ldots,y_{n+m}$ and $y_1,\ldots,y_n$, respectively, it follows that

$$\mathfrak{A}\models_{Y'\cup Z'}\bigwedge_{1\leq j\leq n+m}=(\bar{z}^j,y_j)\wedge(\theta_1\vee\theta_2),$$

and finally

$$\mathfrak{A}\models_X \exists y_1\ldots\exists y_n\exists y_{n+1}\ldots\exists y_{n+m}(\bigwedge_{1\leq j\leq n+m}=(\bar{z}^j,y_j)\wedge(\theta_1\vee\theta_2)). \tag{12}$$

For the converse, it is immediate that the formula in (12) logically implies $\psi^*$. Therefore it is logically equivalent to our original formula $\psi$.

The case $\psi:=\phi_1\wedge\phi_2$ is proved analogously. By the induction hypothesis

$$\phi_1\equiv\exists y_1\ldots\exists y_n(\bigwedge_{1\leq j\leq n}=(\bar{z}^j,y_j)\wedge\theta_1),$$

and

$$\phi_2\equiv\exists y_{n+1}\ldots\exists y_{n+m}(\bigwedge_{n+1\leq j\leq n+m}=(\bar{z}^j,y_j)\wedge\theta_2).$$

Now, analogously to above, we get that $\psi$ is equivalent to the formula $\psi^*$

$$\psi^*:=\exists y_1\ldots\exists y_n\exists y_{n+1}\ldots\exists y_{n+m}((\bigwedge_{1\leq j\leq n}=(\bar{z}^j,y_j)\wedge\theta_1)\wedge(\bigwedge_{n+1\leq j\leq n+m}=(\bar{z}^j,y_j)\wedge\theta_2)).$$

Here we use successively clause 2 of Lemma 2.14. Now the quantifier-free part of $\psi^*$ can be directly transformed to the desired form. $\square$

It is important to note that the width of the dependence atoms in $\psi$ do not change in the transformation of Lemma 3.2. We are now ready to characterize the fragment of ESO corresponding to the logic $\mathcal{D}(k-dep)$.

**Theorem 3.3.** $\mathcal{D}(k-dep)=\mathrm{ESO}_f(k\text{-ary})=\mathrm{ESO}_f^1(k\text{-ary})$.

*Proof.* We show first that $\mathcal{D}(k-dep)\leq\mathrm{ESO}_f(k\text{-ary})$. We prove this by successive transformations on $\phi\in\mathcal{D}(k-dep)$. By case 5 of Lemma 2.14 we can transform $\phi$ into prenex normal form. In this transformation, we might have to replace some bound variables by new ones, but the width of the dependence atoms do not change in the transformation. So $\phi$ is equivalent to a sentence of the form

$$Q^1 x_1\ldots Q^m x_m\theta, \tag{13}$$

where $Q^i\in\{\exists,\forall\}$ and $\theta$ is a quantifier-free formula. We may further assume that all the dependence atomic subformulas of $\theta$ are of the form $=(z_1,\ldots,z_p)$ for some pairwise distinct variables $z_1,\ldots,z_p$; if $\theta$ has a subformula $=(t_1,\ldots,t_p)$, we may pass on to the sentence

$$Q^1 x_1\ldots Q^m x_m\exists z_1\ldots\exists z_p(\bigwedge_{1\leq i\leq p}z_i=t_i\wedge\theta(z_1/t_1,\ldots,z_p/t_p)).$$

Hence we may assume that the sentence (13) satisfies this assumption. Note that, to ensure this property, only new existentially quantified variables need at worst to be introduced. Next we use Lemma 3.2 to transform $\theta$ into an equivalent form

$$\exists y_1 \ldots \exists y_n ( \bigwedge_{1 \leq j \leq n} =(\bar{z}^j, y_j) \wedge \theta^*),$$

where $\theta^*$ is a quantifier-free formula without dependence atoms. Using 6 of Lemma 2.14, it follows that $\phi$ is equivalent to the sentence $\phi'$

$$\phi' := Q^1 x_1 \ldots Q^m x_m \exists y_1 \ldots \exists y_n ( \bigwedge_{1 \leq j \leq n} =(\bar{z}^j, y_j) \wedge \theta^*), \tag{14}$$

where each variable in the tuple $\bar{z}^j$ is in $\{x_1, \ldots, x_m\}$. Proposition 3.1 now implies that $\phi'$ is equivalent to the $\mathrm{ESO}_f(k\text{-ary})$-sentence $\chi$

$$\chi := \exists f_1 \ldots \exists f_n Q^1 x_1 \ldots Q^m x_m \theta', \tag{15}$$

where $\theta'$ is obtained from $\theta^*$ by replacing every occurrence of $y_i$ by the term $f_i(\bar{z}^i)$.

Let us then show that $\mathrm{ESO}_f(k\text{-ary}) \leq \mathrm{ESO}_f^1(k\text{-ary}) \leq \mathscr{D}(k-dep)$. We will show that every sentence $\psi \in \mathrm{ESO}_f(k\text{-ary})$ can be transformed into an equivalent sentence resembling $\chi \in \mathrm{ESO}_f^1(k\text{-ary})$ in Proposition 3.1. First of all, we may certainly assume that the first-order part of $\psi$ is in prenex normal form, i.e.,

$$\psi := \exists f_1 \ldots \exists f_n Q^1 x_1 \ldots Q^m x_m \theta, \tag{16}$$

where $\theta$ is a quantifier-free formula. We still need to make sure that,

($\star$) for each symbol $f_i$, there exists $i_1, \ldots, i_j$, pairwise distinct, such that all terms and subterms in $\theta$ with $f_i$ as the outermost symbol are of the form $f_i(x_{i_1}, \ldots, x_{i_j})$.

This can be accomplished analogously to Theorem 6.15 in [19]. One by one, we replace each occurrence of every term $f(t_1, \ldots, t_j)$ in $\theta$ by a new term $f(z_1, \ldots, z_j)$, where $z_1, \ldots, z_j$ is a fresh tuple of pairwise distinct variables and use the equivalence of $\theta(f(t_1, \ldots, t_j))$ and

$$\forall z_1 \ldots \forall z_j ( \bigwedge_{1 \leq p \leq j} z_p = t_p \to \theta(f(z_1, \ldots, z_j))).$$

In this way, $\theta$ is transformed to an equivalent formula of the form $\forall \bar{z} \theta'$ such that $\theta'$ contains only simple terms of the form $f(z_1, \ldots, z_j)$. If now $\theta'$ contains two occurrences $f_i(z_1^0, \ldots, z_j^0)$ and $f_i(z_1^1, \ldots, z_j^1)$ of the same $f_i$, for $1 \leq i \leq n$, but different variables, we replace the occurrence $f_i(z_1^1, \ldots, z_j^1)$ by $f_i'(z_1^1, \ldots, z_j^1)$ and use the fact that $\forall \bar{z} \theta'(f_i(z_1^0, \ldots, z_j^0), f_i(z_1^1, \ldots, z_j^1))$ is equivalent to

$$\exists f_i' \forall \bar{z} ((( \bigwedge_{1 \leq p \leq j} z_p^0 = z_p^1) \to f_i(z_1^0, \ldots, z_j^0) = f_i'(z_1^1, \ldots, z_j^1)) \wedge \theta'(f_i(z_1^0, \ldots, z_j^0), f_i'(z_1^1, \ldots, z_j^1))).$$

After these transformations, we have translated $\theta$ into the form $\exists \overline{f'} \forall \bar{z} \theta^*$ satisfying ($\star$). Therefore, $\psi$ in (16) is now equivalent to the formula

$$\exists f_1 \ldots \exists f_n Q^1 x_1 \ldots Q^m x_m \exists \overline{f'} \forall \bar{z} \theta^*. \tag{17}$$

Since the functions $f_i'$ are forced to be equal to one of the functions $f_1, \ldots, f_n$, we get that (17) is equivalent to

$$\exists f_1 \ldots \exists f_n \exists \overline{f'} Q^1 x_1 \ldots Q^m x_m \forall \bar{z} \theta^*. \tag{18}$$

The sentence (18) is contained in $\mathrm{ESO}_f^1(k\text{-ary})$. Furthermore, since it satisfies ($\star$), we can directly translate it to the logic $\mathscr{D}(k-dep)$ by Proposition 3.1. $\square$

## 3.1 The logics $\mathscr{D}(k-\forall)$

In this section we consider the logics $\mathscr{D}(k-\forall)$. The case $k=0$ is solved by Proposition 2.13, hence we assume that $k \geq 1$.

We first note that allowing reusing of variables in this context would trivialize the situation. Denote by $\mathscr{D}^*(k-\forall)$ the analogue of $\mathscr{D}(k-\forall)$ in which reusing of variables is allowed. Formally, we let $\mathscr{D}^*(k-\forall)$ denote the class of NNF sentences of $\mathscr{D}$ in which the variables $x_1,\ldots,x_k$ can be universally quantified and the variables $y_i$, $i \in \mathbb{N}$, existentially quantified.

**Proposition 3.4.** $\mathscr{D}^*(1-\forall) = \mathscr{D}$.

*Proof.* Let us assume that the formulas of $\mathscr{D}$ are built from variables $y_i$, where $i \in \mathbb{N}$. We will show that every formula $\phi \in \mathscr{D}$ is equivalent to a formula $\phi^* \in \mathscr{D}^*(1-\forall)$, in which the variable $x$ is quantified universally and the variables $y_i$ are only quantified existentially. We define $\phi^*$ inductive as follows: if $\phi$ is atomic or negated atomic, $\phi^* := \phi$. The connectives $\wedge, \vee$, and $\exists$ are also translated in the obvious way. If $\phi$ is of the form $\forall y_i \psi$, we define $\phi^*$ as
$$\phi^* := \forall x \exists y_i (x = y_i \wedge \psi^*).$$
Note that, by the construction, $x$ does not appear free in $\psi^*$ so $\phi$ and $\phi^*$ have the same free variables. It is now easy to prove using induction on the complexity of $\phi$ that for all models $\mathfrak{A}$ and teams $X$
$$\mathfrak{A} \models_X \phi \iff \mathfrak{A} \models_X \phi^*.$$
□

Our goal is now to characterize the fragments of ESO corresponding to the logics $\mathscr{D}(k-\forall)$. We first discuss some results regarding the relevant fragments of ESO.

**Proposition 3.5.** $\mathrm{ESO}_f^1(k\forall) \leq \mathrm{ESO}_f^1(k\forall, \exists^*) \leq \mathrm{ESO}_f(k\forall)$.

*Proof.* The first inequality holds by definition. The second inequality can be proved by noting that the existential quantifiers $Q^i$ can be replaced by Skolem functions and the resulting sentence will be in the logic $\mathrm{ESO}_f(k\forall)$. □

Contrasting with the result of Theorem 3.3, it is not known whether the classes $\mathrm{ESO}_f^1(k\forall)$ and $\mathrm{ESO}_f(k\forall)$ are equal. However, the next proposition shows that the gap between these two logics is not that big after all.

**Proposition 3.6.** *For every sentence $\phi$ in $\mathrm{ESO}_f(k\forall)$ there is a sentence $\phi^*$ in $\mathrm{ESO}_f^1(2k\forall)$ such that, for all $\mathfrak{A}$:*
$$\mathfrak{A} \models \phi \iff \mathfrak{A} \models \phi^*.$$

*Proof.* Assume that $\phi$ is of the form
$$\exists f_1 \ldots \exists f_n \forall x_1 \ldots \forall x_k \psi$$
where $\psi$ is quantifier free. Let us call $so(\phi)$ the set of existentially quantified function symbols that appear in $\phi$. By Theorem 2.18 we can also suppose that the functions of $so(\phi)$ are of arity $k$. By introducing new function symbols, we will normalize the formula $\phi$ step by step.

1. For every term $f(t_1,\ldots,t_m)$ in $\phi$, which is not of the form $f(x_1,\ldots,x_k)$ or $f(g_1(\bar{x}),\ldots,g_m(\bar{x}))$, where $\bar{x} = (x_1,\ldots,x_k)$, $g_i \in so(\phi)$ for $1 \leq i \leq m$, and $f, g_1,\ldots,g_m$ are pairwise distinct, we introduce new functions $h_1,\ldots,h_m$ not in $so(\phi)$ and replace all occurrences of $f(t_1,\ldots,t_m)$ by $f(h_1(\bar{x}),\ldots,h_m(\bar{x}))$ using the equivalence:

$$\models \forall \bar{x}\ \psi(f(t_1,\ldots,t_m)) \leftrightarrow \forall \bar{x}\ \psi(f(h_1(\bar{x}),\ldots,h_m(\bar{x}))) \wedge \bigwedge_{j=1}^{m} h_j(\bar{x}) = t_j.$$

   After these transformations, the composition depth of all terms is bounded by 2.

2. Transform $\phi$ in such a way that no function symbol appears both as an inner and an outer function symbol even in different composed terms. This can be done by systematically renaming all inner terms $g_i(\bar{x})$ by a new term $h_i(\bar{x})$ with $h_i \notin so(\phi)$ and using the equivalence:

$$\models \forall \bar{x}\ \psi(f(g_1(\bar{x}),\ldots,g_m(\bar{x}))) \leftrightarrow \forall \bar{x}\ \psi(f(h_1(\bar{x}),\ldots,h_m(\bar{x}))) \wedge \bigwedge_{i=1}^{m} g_i(\bar{x}) = h_i(\bar{x}).$$

3. Finally, for convenience, one forces that for each function $f \in so(\phi)$, there is at least one occurrence of the term $f(\bar{x})$. For that, it suffices to add in conjunction with the formula a dummy equality $f(\bar{x}) = h(\bar{x})$ where $h \notin so(\phi)$.

We are now ready to make the final transformation on $\phi$. Note that those $f \in so(\phi)$ that only appear as an inner symbol in composed terms, have only occurrences of the form $f(x_1,\ldots,x_k)$ in $\phi$. Therefore, it suffices to consider $f \in so(\phi)$ having at least one occurrence of the form $f(g_1(\bar{x}),\ldots,g_k(\bar{x}))$ in $\phi$. Let $term(f)$ be the set of terms involving $f$. The elements $\tau_1(f),\ldots,\tau_{m_f}(f)$ of $term(f)$ are of the form

$$\tau_i(f) = f(g_{i,1}(\bar{x}),\ldots,g_{i,k}(\bar{x}))$$

where all functions $g_{i,j}$ are in $(so(\phi) \cup \{pr_1,\ldots,pr_k\}) \setminus \{f\}$ where each $pr_j$ is the projection function on the $j$th argument. Without loss of generality, we may suppose that $\tau_1(f) = f(x_1,\ldots,x_k)$. Let us introduce new functions symbols $h_1,\ldots,h_{m_f}$ not yet in $so(\phi)$. Then, the following equivalence holds:

$$\models \forall \bar{x}\psi \leftrightarrow \forall \bar{x} \forall \bar{x}'\ \bigwedge_{f \in so(\phi)} \bigwedge_{i>1}^{m_f} (x_1' = g_{i,1}(\bar{x}) \wedge \cdots \wedge x_k' = g_{i,k}(\bar{x})) \to f(\bar{x}') = h_i(\bar{x})$$
$$\wedge (\bar{x} = \bar{x}' \to \psi^*)$$

where $\psi^*$ is obtained from $\psi$ by replacing every occurrence of $\tau_1(f)$ by $f(\bar{x}')$ and every occurrence of $\tau_i(f)$, for $i > 1$, by $h_i(\bar{x})$. Note that in the original $\phi$, all simple terms involving $f$ (i.e., not as an outermost symbol in a composition) are all already of the form $f(\bar{x})$ (from the normalization process). They can then be transformed into $f(\bar{x}')$ directly. This step is repeated for every function of $so(\phi)$ appearing as an outer symbol in a composed term. $\square$

Now we turn to the characterization of the logics $\mathscr{D}(k - \forall)$. The following lemma will be used.

**Lemma 3.7.** *Every sentence $\phi \in \mathscr{D}(k - \forall)$ is equivalent to a sentence $\phi^* \in \mathscr{D}(k - \forall)$ which is in prenex normal form.*

*Proof.* The equivalences 1-4 of Lemma 2.14 can be applied to transform $\phi$ into prenex normal form since each variable in $\phi$ is quantified exactly once. $\square$

We are now ready for the main result of this section.

**Theorem 3.8.** $\mathscr{D}(k-\forall) = \mathrm{ESO}_f^1(k\forall, \exists^*)$.

*Proof.* Note first that $\mathrm{ESO}_f^1(k\forall, \exists^*) \leq \mathscr{D}(k-\forall)$ follows immediately by Proposition 3.1. We will now show that also the converse holds. Let $\phi \in \mathscr{D}(k-\forall)$ be a sentence. We will construct a sentence $\chi \in \mathrm{ESO}_f^1(k\forall, \exists^*)$ equivalent to $\phi$. We use the same idea as in the proof of Theorem 3.3. By Lemma 3.7, we may assume that $\phi$ is in prenex normal form, i.e., $\phi$ is of the form

$$Q^1 x_1 \ldots Q^m x_m \theta,$$

where $Q^i \in \{\exists, \forall\}$ and $\theta$ is a quantifier-free formula. Analogously to the proof of Theorem 3.3, we may also assume that all the dependence atomic subformulas of $\theta$ are of the form $=(z_1, \ldots, z_p)$ for some pairwise distinct variables $z_1, \ldots, z_p$. As remarked in the proof of Theorem 3.3, the cost of assuming this property is the introduction of new existentially quantified variables (i.e., the number of universal quantifiers does not change). Next we use Lemma 3.2 to transform $\theta$ into an equivalent form

$$\exists y_1 \ldots \exists y_n (\bigwedge_{1 \leq j \leq n} =(\bar{z}^j, y_j) \wedge \theta^*),$$

where $\theta^*$ is a quantifier-free formula without dependence atoms. By 6 of Lemma 2.14, we now get that $\phi$ is equivalent to the sentence $\phi'$

$$Q^1 x_1 \ldots Q^m x_m \exists y_1 \ldots \exists y_n (\bigwedge_{1 \leq j \leq n} =(\bar{z}^j, y_j) \wedge \theta^*), \tag{19}$$

where each variable in the tuple $\bar{z}^j$ is in $\{x_1, \ldots, x_m\}$ and the variables in $\bar{z}^j$ are pairwise distinct. Note also that the quantifier prefix $Q^1 x_1 \ldots Q^m x_m$ has at most $k$ universal quantifiers. Proposition 3.1 now implies that $\phi'$ is equivalent to a sentence $\chi \in \mathrm{ESO}_f^1(k\forall, \exists^*)$. □

**Corollary 3.9.** *For* $k \in \mathbb{N}^*$, $\mathscr{D}(k-\forall) \leq \mathscr{D}(k-dep)$.

*Proof.* The claim follows by the following chain of inequalities:

$$\mathscr{D}(k-\forall) \leq \mathrm{ESO}_f^1(k\forall, \exists^*) \leq \mathrm{ESO}_f(k\forall) \leq \mathrm{ESO}_f(k\text{-ary}, k\forall) \leq \mathrm{ESO}_f(k\text{-ary}) \leq \mathscr{D}(k-dep),$$

where the first three inequalities hold by Theorem 3.8, Proposition 3.5, Theorem 2.18, and the last by Theorem 3.3. □

## 4 Hierarchy theorems for $\mathscr{D}$

In this section we use the results of the previous sections to show expressibility hierarchies for fragments of dependence logic. From the results of the preceding sections, one obtains the following result.

**Corollary 4.1.** *For every* $k \in \mathbb{N}^*$, *on every signature:* $\mathscr{D}(k-\forall) < \mathscr{D}(k+1-dep)$.

*Proof.* By Theorem 3.8 and Proposition 3.5, it holds that $\mathscr{D}(k-\forall) \leq \mathrm{ESO}_f(k\forall)$. The following chain of inclusions now hold using Corollary 2.21 and 2.18:

$$\mathrm{ESO}_f(k\forall) = \mathrm{NTIME}_{\mathrm{RAM}}(n^k) < \mathrm{NTIME}_{\mathrm{RAM}}(n^{k+1}) \leq \mathrm{ESO}_f(k+1\text{-ary}) = \mathscr{D}(k+1-dep).$$

□

A hierarchy result for $\mathscr{D}(k-\forall)$ can also be stated.

**Corollary 4.2.** *For every $k \in \mathbb{N}^*$, on every signature: $\mathscr{D}(k-\forall) < \mathscr{D}(2k+2-\forall)$.*

*Proof.* Again, using Theorem 3.8 and Proposition 3.5, it holds that

$$\mathrm{ESO}_f^1(k\forall) \leq \mathscr{D}(k-\forall) \leq \mathrm{ESO}_f(k\forall).$$

By the "hierarchy" Corollary 2.21, $\mathrm{ESO}_f(k\forall) < \mathrm{ESO}_f(k+1\forall)$ and from Proposition 3.6, it holds:

$$\mathrm{ESO}_f(k+1\forall) \leq \mathrm{ESO}_f^1(2k+2\forall).$$

Then, the result follows by applying Theorem 3.8 again. □

The above hierarchy on $\mathscr{D}(k-\forall)$ is not tight. However, the following is easily seen to be true (in the results that follow, that do not hold for all signatures or every value of $k$, we make explicit the parameter $\tau$ in the notation).

**Corollary 4.3.** *For every signature $\tau$, there exists an infinity of $k \in \mathbb{N}$ such that $\mathscr{D}(k-\forall)[\tau] < \mathscr{D}(k+1-\forall)[\tau]$.*

*Proof.* By Corollary 4.2, it holds that $\mathscr{D}(k-\forall) < \mathscr{D}(2k+2-\forall)$ for all $k \geq 1$ and all signatures $\tau$. Fix $k \in \mathbb{N}^*$, the result above implies that there exists $h \in (k, 2k+1)$ such that $\mathscr{D}(h-\forall)[\tau] < \mathscr{D}(h+1-\forall)[\tau]$. Since the number of pairwise disjoint intervals of the form $(k, 2k+1)$ is infinite, the result follows. □

We now turn back to logics $\mathscr{D}(k-dep)$. The time hierarchy theorem can be used in the context of the logics $\mathrm{ESO}_f(k\forall)$ but not directly with the logics $\mathrm{ESO}_f(k$-ary) or $\mathrm{ESO}(k$-ary). Ajtai [1] showed that also the logics $\mathrm{ESO}(k$-ary) form a strict hierarchy with respect to $k$ if the signature is allowed to vary. By an easy reduction, one can improve the result to show that it is also the case for the logics $\mathrm{ESO}_f(k$-ary) (see [3] for the separation of the two first levels). To summarize:

**Theorem 4.4** ([1])**.** *Let $R$ be a $k+1$-ary relation symbol. Then the property "$|R|$ even" cannot be defined in the logic $\mathrm{ESO}_f(k$-ary).*

Since "$|R|$ even" is expressible in $\mathrm{ESO}(k+1$-ary), we get that, for all $k$

$$\mathrm{ESO}_f(k\text{-ary})[\tau_{k+1}] < \mathrm{ESO}(k+1\text{-ary})[\tau_{k+1}],$$

where $\tau_{k+1} = \{R\}$ and $R$ is $k+1$-ary. By Theorem 3.3, the logics $\mathscr{D}(k-dep)$ also form a hierarchy with respect to expressive power using Theorem 4.4.

**Theorem 4.5.** *Let $k \geq 1$ and $\tau_{k+1} = \{R\}$ where $R$ is $k+1$-ary. Then $\mathscr{D}(k-dep)[\tau_{k+1}] < \mathscr{D}(k+1-dep)[\tau_{k+1}]$. In particular, the property "$|R|$ even" is definable in $\mathscr{D}(k+1-dep)[\tau_{k+1}]$ but not in $\mathscr{D}(k-dep)[\tau_{k+1}]$.*

*Proof.* The claim directly follows from Theorems 3.3 and 4.4. □

The above result provides a kind of "subdiagonal" hierarchy when the maximal arity of a relation in the signature is greater than the authorized arity of a dependence atomic formula. Could a better result be proved? In particular, is it true that for every signature $\tau$ and all $k$, $\mathscr{D}(k-dep)[\tau] < \mathscr{D}(k+1-dep)[\tau]$? By Theorem 3.3, such a result would imply that $\text{ESO}_f(k\text{-ary}) < \text{ESO}_f(k+1\text{-ary})$ for every signature. Although it is reasonably conjectured to be true, such a result is not yet known. In the particular case of $\tau = \emptyset$, it would imply that there exist sets of integers definable by first-order sentences (i.e., which are spectra of first-order sentences) with predicates of maximal arity $k+1$ which are not definable by sentences with predicates of arity $k$. This latter question is left open in [5] (it concerns the so-called Spectrum Arity Hierarchy, see also [6]) and has not received a satisfiable answer since then despite numerous efforts. Proving an equivalent hierarchy for $\mathscr{D}(k-dep)$ is a challenging and difficult task with consequences to fields beyond dependence logic.

## Conclusion

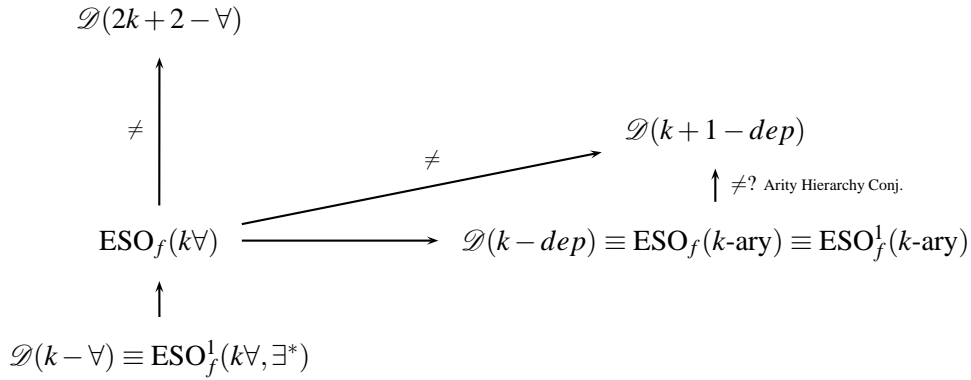

Figure 1: Summary of inclusions (for all signatures and all $k \geq 1$)

We have pinned down the fragments of ESO corresponding to the fragments $\mathscr{D}(k-\forall)$ and $\mathscr{D}(k-dep)$ of $\mathscr{D}$ (Figure 1 summarizes the main relationships between logics considered in this paper). Our results explain how important syntactic parameters, the maximal width of dependence atoms, and the number of universal quantifiers in a sentence, reflect on its data complexity. We also showed that fixing either of the parameters will lead to a loss in expressive power. The following questions remain open.

1. Is it the case that $\mathscr{D}(k-\forall) < \mathscr{D}(k+1-\forall)$ for all $k$ *and* all signatures?

2. Does $\mathscr{D}(k-\forall)[\tau] < \mathscr{D}(k-dep)[\tau]$ hold for all signatures $\tau$?

3. Is there a signature $\tau$ for which $\mathscr{D}(k-dep)[\tau] < \mathscr{D}(k+1-dep)[\tau]$ holds for all $k$?

Remark that Corollary 4.3 does not answer Question 1 since it shows that for every signature the inclusion is strict for inifinitely many $k$ but not for all $k$ yet. It is worth noting that, by Lemma 2.20, 2 holds if $\tau$ has arity greater than $k$. Also, 3 is open already in the case $\tau = \emptyset$.